\documentclass[twocolumn,aps,prb,showpacs]{revtex4}
\usepackage{graphicx}
\usepackage{amsfonts}
\usepackage{bm}

\begin{document}
\title{Spin-orbit coupled Bose-Einstein condensates}
\author{Tudor D. Stanescu, Brandon Anderson  and Victor Galitski }
\affiliation{Department of Physics and Joint Quantum Institute,
University of Maryland, College Park, MD 20742-4111}

\begin{abstract}
We consider a many-body system of pseudo-spin-$1/2$ bosons with spin-orbit interactions, which couple the momentum and the internal pseudo-spin degree of freedom created by spatially varying laser fields. The corresponding single-particle spectrum is generally anisotropic and contains two degenerate minima at finite momenta. At low temperatures, the many-body system condenses into these minima generating a new type of entangled Bose-Einstein condensate. We show that in the presence of weak density-density interactions the many-body ground state is characterized by a twofold degeneracy. The corresponding many-body wave function describes a condensate of ``left-'' and ``right-moving'' bosons. By fine-tuning the parameters of the laser field, one can obtain a bosonic version of the spin-orbit coupled Rashba model. In this symmetric case, the degeneracy of the ground state is very large, which may lead to phases with nontrivial topological properties. We argue that the predicted new type of Bose-Einstein condensates can be observed experimentally via time-of-flight imaging, which will show characteristic multipeak structures in momentum distribution.
\end{abstract}

\maketitle

\section{Introduction} \label{I}

Bose-Einstein condensation is an old and thoroughly studied
quantum phenomenon, where a many-body system of bosons undergoes a
phase transition in which a single-particle state becomes
macroscopically occupied. This phenomenon has been observed in
condensed matter systems and more recently in experiments on cold
atomic gases,\cite{Leggett,rmp.BEC} which provided a unique
avenue to visualize the formation of the Bose-Einstein condensate
(BEC). Bose-Einstein condensation is a phase transition driven
mostly by the statistics of the underlying bosonic excitations and
not by interactions. The statistics of basic particles are
determined by the particle spin via the fundamental Pauli
spin-statistics theorem:\cite{Pauli}  The spin must be integer for
bosons and half-integer for fermions.

In this paper, we discuss a cold atomic system of multi-level
bosons moving in the presence of spatially-varying laser fields,
which give rise to an emergent pseudo-spin-$1/2$  degree of
freedom for the bosons. We emphasize from the outset that the
symmetry operations in the pseudo-spin space are not related to
real-space rotations and thus there is no contradiction between
the existence of the pseudo-spin-$1/2$ bosons and the fundamental
Pauli theorem. To ``create'' the pseudo-spin-$1/2$ bosons, one can
use the experimental setup, proposed in
Refs.~\onlinecite{oberg1,Jaksch,Rusec,Osterloh,Zhu,Clark,SZG}, in which
three degenerate hyperfine ground states $|1\rangle$, $|2\rangle$,
$|3\rangle$ are coupled to an excited state $|0\rangle$ by
spatially varying laser fields. This ``tripod scheme'' leads to
the appearance of a pair of degenerate dark states, spanning a
subspace which is well separated in energy from two nondegenerate
bright states. The coupling between the dark and the bright states
is very weak and will be neglected (adiabatic approximation). The
parameter which labels the dark states plays the role of a
pseudo-spin index. This emergent pseudo-spin degree of freedom is
similar to that studied recently in the context of spinor
condensates.\cite{SC,Leggett1/2} In particular, various aspects of the 
pseudo spin-$1/2$ boson physics were 
addressed\cite{SBEC1,SBEC2,SBEC3,SBEC4,SBEC5,SBEC6,SBEC7,SBEC8,SBEC9,SBEC10} 
using two hyperfine 
states to support the internal degree of freedom associated with the 
pseudo spin. The key distinctive feature of the systems studied
in this paper is that the single-particle Hamiltonian projected
onto the sub-space of the degenerate dark states generally
possesses a non-Abelian gauge structure. I.e., the kinetic term of
the effective Hamiltonian has the form $\check{H}_{\rm kin} =
\left( {\bf p} \check{1} - \check{\bf A} \right)^2/2m$, where
$\check{\bf A}({\bf r})$ is a matrix in the pseudo-spin space. In
a recent Letter,\cite{SZG} we pointed out that under certain
conditions this non-Abelian gauge structure is equivalent to a
spin-orbit interaction. To understand the nature of this
interaction, we note that the dark states are eigenstates of an
atom at rest. Once the atom moves in the spatially modulated laser
field, the dark state label, i.e., the pseudo-spin index, is not a
good quantum number and the pseudo-spin starts to precess about
the direction of the momentum. This coupling between the internal
degree of freedom associated with the dark state subspace and the
orbital movement of the particle represents the spin-orbit
interaction. The spin-orbit coupling parameters can be adjusted by
changing the properties of the spatially modulated light beams.
These conclusions are based entirely on single-particle physics;
the particle statistics play no role. Below, we consider a
many-particle system of bosons within this tripod scheme. Due to
spin-orbit coupling, the degenerate ground states of the system
correspond to non-zero momenta, leading to a new type of BEC, 
the spin-orbit coupled Bose-Einstein condensate (SOBEC).

The article is organized as follows: In Section \ref{II} we
introduce our model and discuss the properties of a non-interacting
many body system of bosons with spin-orbit interactions.  
We find that, in general, the single-particle spectrum is
characterized by two degenerate minima at finite momenta and we
determine the transition temperature for the  bosons
condensing into these minima. In Section \ref{III} we study the
effects of density-density interactions using a generalized
Bogoliubov transformation (Subsection \ref{IIIa}). We show the quasiparticle
excitation spectrum contains an anisotropic free particle
component and an anisotropic sound similar to the conventional
Bogoliubov phonon. By calculating the energy of the condensate, we
find that for a system of $N$ bosons the $(N+1)$-fold degeneracy
of the non-interacting ground state is reduced by the interactions
to a two-fold degeneracy corresponding to ``left-'' or
``right-moving'' particles. The corresponding many-body wavefunction 
describes a NOON state,\cite{NOON} suggesting that future studies of 
the SOBEC state in the context of quantum entanglement and quantum 
interference are highly relevant. 
 For completeness, we also derive the 
 Gross-Pitaevskii equations for the spin-orbit coupled condensate 
(Subsection \ref{IIIb}). Linearizing the coupled non-linear equations 
in the vicinity of a stationary solution leads to a spectrum of 
excitations that reproduces the generalized Bogoliubov result. 
A possible experimental signature of the new type of SOBEC is described
in Section \ref{IV}. We argue that a SOBEC can be observed via 
time-of-flight imaging, which will show a characteristic multi-peak 
structure of the density profiles. We demonstrate that such a measurement 
generates distinct outputs for ``left-'' and ``right-moving'' condensates
and thus can be viewed as a measurement of a qubit. A summary of the paper 
along with our  conclusions are presented in Section \ref{V}.

\section{Spin-orbit interacting Hamiltonian and single-particle spectrum} \label{II}

We start with the following many-body Hamiltonian describing
spin-orbit coupled bosons,
\begin{equation}
{\hat\mathcal{H}}= \sum\limits_{{\bf p};\alpha,\beta}
\hat{b}_{\alpha {\bf p}}^{\dagger} \left\{ {{\bf p}^2 \over 2m}
\check{1} - v p_x \check{\sigma}_2 - v' p_y
\check{\sigma}_3\right\}_{\alpha \beta}  \hat{b}_{\beta {\bf p}}
\label{H},
\end{equation}
where $\hat{b}_{\alpha {\bf p}}^{\dagger}$ and $\hat{b}_{\alpha
{\bf p}}$ are the creation and annihilation operators for bosons
in the state with momentum ${\bf p}$ and pseudo-spin $\alpha =
 {\uparrow,\downarrow}$, $\check{\sigma}_{i}$ are the Pauli
matrices in the pseudo-spin space, and the parameters $v$ and $v'$
characterize the strength and anisotropy of the  spin-orbit
coupling.   We reiterate that this type of spin-orbit-coupled
Hamiltonian (\ref{H}) will appear within the recently proposed
tripod scheme\cite{oberg1,Jaksch,Rusec,Osterloh,Zhu,Clark,SZG} in
which three hyperfine ground states of an atom $\left\vert
1\right\rangle$, $\left\vert 2\right\rangle$, and $\left\vert
3\right\rangle$ are coupled to an excited state $\left\vert
0\right\rangle$ via spatially modulated laser fields. The
underlying laser-atom Hamiltonian is
\begin{equation}
{\mathcal{H}}_{\rm a-l}= \Omega_{0}\left\vert 0\right\rangle
\left\langle 0\right\vert + \sum\limits_{\mu=1}^{3} \left[
\Omega_{\mu}({\bf r}) \left\vert 0\right\rangle \left\langle
\mu\right\vert + {\rm h.~c.} \right],        \label{Ham_al}
\end{equation}
where $\Omega_0$ is a constant detuning to the excited state and
the Rabi frequencies consistent with the realization of an
effective spin-orbit interaction can be taken as $\Omega_{1}({\bf
r})= \Omega\sin\theta\cos(mv_ax)e^{imv_by}$, $\Omega_{2}({\bf
r})=\Omega\sin\theta\sin(mv_ax)e^{imv_by}$, and $\Omega_{3}({\bf
r})=\Omega\cos\theta$, with $\Omega$, $\theta$, $v_a$ and $v_b$
being constants (see, e.g., Refs.~\onlinecite{Rusec}
and~\onlinecite{SZG} for details). Diagonalizing the atom-laser
Hamiltonian (\ref{Ham_al}) via a position-dependent rotation
$R_{\mu\alpha}({\bf r})$, with $\alpha\in\{\uparrow, \downarrow,
b_1, b_2\}$ and $\mu\in\{0, 1, 2, 3\}$, generates a pair of
degenerate dark states
\begin{eqnarray}
\left\vert \uparrow \right\rangle &=& \sin\Phi_x e^{-iS_y}\vert
1\rangle -
\cos\Phi_x e^{-iS_y}\vert 2\rangle,   \label{darkSts} \\
\left\vert \downarrow \right\rangle &=& \cos\theta\cos\Phi_x
e^{-iS_y}\vert 1\rangle +\cos\theta\sin\Phi_x e^{-iS_y}\vert
2\rangle - \sin\theta\vert 3\rangle, \nonumber
\end{eqnarray}
with $\Phi_x = m v_a x$ and $S_y = m v_b y$, and two
non-degenerate bright states $\vert b_{1(2)}\rangle$
The pseudo-spin-$1/2$ structure emerges when the problem is projected 
onto the subspace
spanned by the pair of degenerate dark states.\cite{SZG} Applying
the position-dependent rotation $R_{\mu\alpha}({\bf r})$ to
the kinetic energy term in the Hamiltonian generates a coupling of
the pseudo-spin to momentum [see Eq.~(\ref{H})] with $v = v_a
\cos{\theta}$ and $v^\prime = v_b \sin^2{(\theta/2)}$, in the given
parametrization. These coupling constants can be easily adjusted
by changing the parameters $v_a$, $v_b$ and $\theta$  of the laser fields, 
which provides a
knob to tune the strength and form of the spin-orbit interaction.

\begin{figure}[tbp]
\begin{center}
\includegraphics[width=0.43\textwidth]{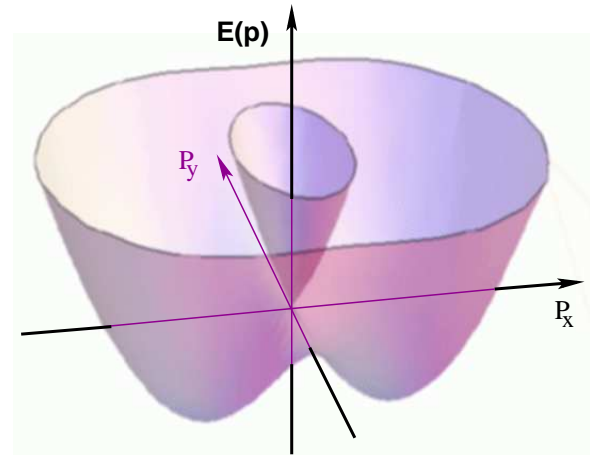}
\end{center}
\caption{(color online):~ Schematic picture of the band structure
described by Eq. (\ref{E}) with $v/v' = 2.5$ for a constant value
of $p_z$. The inside sheet represents the $\lambda = +1$ band,
while the outside sheet corresponds to $\lambda=-1$ and has a
double-well structure with minima at $p_x = \pm m v$ and $p_y =
0$. \vspace*{-0.15in}} \label{Fig1}
\end{figure}

Now, we concentrate on the generic case characterized by
anisotropic spin-orbit interactions and assume for concreteness
that $v > v^\prime>0$. The trap potential and the inter-particle
interaction are initially disregarded and discussed in the
following sections. The single-particle spectrum of
Hamiltonian~(\ref{H}) is (see Fig.~\ref{Fig1}):
\begin{equation}
\label{E} {E}_{\lambda}({\bf p}) = {{\bf p}^2 \over 2m} + \lambda
\sqrt{ v^2 p_x^2 + v'^2 p_y^2}, 
\end{equation}
where $\lambda=\pm 1$ labels the bands.
The corresponding eigenfunctions $\vec{\phi}_{\lambda{\bf p}}({\bf
r}) =  e^{i{\bf p r}} \vec{U}_{\lambda}(\chi_{\bf p})$ are spinors with
components
\begin{equation}
\label{u.up} {U}_{\uparrow \lambda}(\chi_{\bf p}) = {\left[
\sqrt{\cos^2{\chi_{\bf p}} + \Delta^2 \sin^2{\chi_{\bf p}}} - \Delta \lambda
\sin{\chi_{\bf p}} \right]^{1/2} \over \sqrt{2} \left[ \cos^2{\chi_{\bf p}} +
\Delta^2 \sin^2{\chi_{\bf p}} \right]^{1/4}}\,\,\,\,\,\mbox{ and}
\end{equation}
\begin{equation}
\label{u.down} {U}_{\downarrow \lambda}(\chi_{\bf p}) = -i\lambda~
\mbox{sign}\,[\cos{\chi_{\bf p}}]~{U}_{\uparrow -\lambda}(\chi_{\bf p}),
\end{equation}
where $\chi_{\bf p}$ is the azimuthal angle in the $(p_x,p_y)$-plane and
$\Delta = v^\prime/v <1$. The unitary matrix $U_{\alpha \lambda} (\chi_{\bf p})$
diagonalizes the Hamiltonian~(\ref{H}) (where $\alpha =
\uparrow,\downarrow$ corresponds to the pseudo-spin index and
$\lambda = \pm 1$ labels the eigenstates). It is obvious from
Eq.~(\ref{E}) that the spectrum of the single particle problem
contains two minima at $\lambda = -1$ and momenta $p_y = p_z = 0$
and $p_x = \pm mv \ne 0$ (see Fig.~\ref{Fig1}). Consequently, the
single particle ground-state is double-degenerate and  the most
general expression for the corresponding wave-function is
\begin{equation}
\Psi_{\rm dw}({\bf r}) = \sqrt{w_{\rm L}}\left(\begin{array}{cc} 1
\\-i\end{array}\right)e^{-imvx + i\phi_{\rm L}}
+  \sqrt{w_{\rm R}}\left(\begin{array}{cc}
1\\i\end{array}\right)e^{imvx + i\phi_{\rm R}},    \label{Psi.dw}
\end{equation}
where $w_{\rm L} \ge 0$ and $w_{\rm R} \ge 0$ are the  fractions
of ``left-'' and ``right-moving'' states subjected to the
constraint $w_{\rm L} + w_{\rm R} = 1$, while $\phi_{\rm L}$ and
$\phi_{\rm R}$ are arbitrary phases. Note that by
left/right-moving states we mean states with non-zero momentum
average, $\langle {\bf p}\rangle =\mp mv {\bf e}_x$. However, the
corresponding average velocity vanishes $\langle {\bm \nabla}_{\bf
p} \check{\cal H}({\bf p})\rangle = {\bf 0}$, so that
quasiparticles characterized by these non-zero momentum
single-particle states are not actually ``moving'', as long as the
laser fields generating the spin-orbit coupling are maintained.
Note that rotations in the manifold of the double-well
ground-states are distinct from rotations in the pseudo-spin
Hilbert space, as real-space and pseudo-spin coordinates are mixed
up by the spin-orbit interaction. The two-fold degeneracy of the
single-particle ground state is preserved if the system is placed
in a harmonic trap. For a potential  $V_{\rm trap} = m \omega^2
{\bf r}^2 /2$, we can write the Sch{\"o}dinger equation in
momentum representation: The trap potential plays the role of
``the kinetic energy'' and the real kinetic term produces a
double-well potential in momentum space, see Fig.~\ref{Fig1}. The
tunnelling processes connect the degenerate vacua in momentum
space~\cite{Polyakov}. However, they do not eliminate the
double-degeneracy of the single-particle states, which is
protected by the Kramers-like symmetry (see Section \ref{IIIb}).

At low temperatures, the many-body Bose system (\ref{H}) condenses
into the single-particle states corresponding to the double-well
minima. The transition temperature of this double-well SOBEC can be
calculated using standard text-book procedures.\cite{AGD}  Let us
assume that near and below the transition the band with $\lambda =
+1$ does not contribute and that we can expand the spectrum near
the minima of the band (\ref{E}). We define the momentum ${\bf q}$
in the vicinity of the left/right minima as follows: ${\bf p} =
\pm mv {\bf e}_x + {\bf q}$, with $q \ll mv$. Eq.~(\ref{E}) leads
to the anisotropic spectrum: 
\begin{equation}
\delta E({\bf q}) = \frac{q_x^2 + q_z^2}{2m} + \left[1-\left(\frac{v^{\prime}}{v}\right)^2\right]\frac{q_y^2}{2 m}.  \label{delE}
\end{equation}
The transition temperature is
\begin{equation}
T_{\rm c} = \frac{\pi}{2} \left[ \frac{4}{\zeta(3/2)}\right]^{\frac32} \left[
1 - \left( \frac{v^\prime}{v} \right)^2 \right]^{\frac13} \frac{n^{\frac23}}{m}.
\label{Tc}
\end{equation}
We see that  if $n^{1/3} \left[ 1 - \left( {v' / v} \right)^2
\right]^{1/6}\ll mv$, our approximation is justified and, in
particular, the density of particles in the upper band $\lambda =
+1$ is exponentially small. 

\begin{figure}[tbp]
\begin{center}
\includegraphics[width=0.41\textwidth]{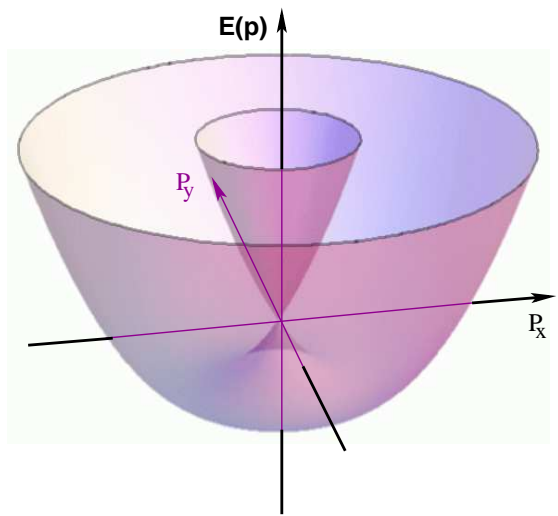}
\end{center}
\caption{(color online):~ Schematic picture of the band structure described by Eq. (\ref{E}) for the isotropic Rashba-type case with $v= v^\prime$ for $p_z=0$. The inside sheet represents the $\lambda = +1$ band, while the outside sheet corresponds to $\lambda=-1$ and has  minima a one-dimensional circle $\sqrt{p_x^2 +p_y^2} = mv$.} \label{Fig1B}
\end{figure}

In the isotropic limit
$\Delta=v^{\prime}/v\rightarrow 1$, the transition temperature formally
vanishes. Note that in the isotropic case $v=v'$ the spin-orbit
term of the Hamiltonian (\ref{H}) is equivalent to the Rashba
model~\cite{RD} and can be reduced to the latter via the rotation
$\exp{(i \pi \check{\sigma}_2/4)}$ in the pseudo-spin space. In
this case, the spectrum (\ref{E}) has  minima on a one-dimensional
circle $\sqrt{p_x^2 +p_y^2} = mv$ (see Fig. \ref{Fig1B}). 
The single-particle ground-state is infinitely degenerate and the most general
expression for the corresponding wave-function is
\begin{equation}
\Psi_{\rm ring}({\bf r}) =\int\limits_0^{2 \pi} {d \chi \over 2
\pi} \sqrt{w(\chi)} ~\vec{U}_-(\chi) e^{i \phi(\chi)} e^{\left[
imv ( x \cos{\chi} + y \sin{\chi}) \right]},  \label{pri_ring}
\end{equation}
where $w(\chi) > 0$ is the angle-dependent weight of the
Bose-condensate on a circle [$\int d\chi/(2\pi) w(\chi) = 1$] and
$\phi(\chi)$ is the angle-dependent phase. An especially
interesting class of ground states corresponds to $w(\chi)$  not
vanishing anywhere on the circle. In this case, the phase
$\phi(\chi)$ must satisfy the constraint $\phi(\chi + 2 \pi) -
\phi(\chi)= 2 \pi n$, with $n \in \mathbb{Z} =\pi_1(S^1)$ being an
integer winding number. Therefore, there may exist a number of
topologically distinct ground states (characterized by the winding
number), which can not be deformed into one another via any
continuous transformation. We note here that a transition into the
ring SOBEC is similar to a ``weak-crystallization transition''
discussed by Brazovsky~\cite{Braz} (see also,
Refs. \onlinecite{ala-Braz}). In this case, the phase volume of
fluctuations is very large, which drives the (classical)
transition first order. Even though the transition temperature
into the ring SOBEC vanishes in the thermodynamic limit, in a finite
trapped system, the energy scale for the crossover into this state
will be non-zero.\cite{RMP.BEC1}

\section{Effects of density-density interaction} \label{III}

The most general  ground-state many-body wave-function of a
non-interacting ``double well BEC'' is
\begin{equation}
|| \Psi_{N} \rangle = \sum_{n=0}^{N} \frac{c_n}{\sqrt{n!(N-n)!}}~
\left(\hat{B}_L^\dagger\right)^n\left(\hat{B}_R^\dagger\right)^{N-n}
||{\rm vac}\rangle, \label{PsiNn}
\end{equation}
where $n$ and $N-n$ are the numbers of ``left-'' and
``right-movers,''  $\hat{B}_{L/R}^\dagger$ are the corresponding
creation operators, and $c_n$ are arbitrary coefficients
satisfying $\sum_n~|c_n|^2 = 1$. In the absence of spin-orbit
interaction, a two-component bosonic system has a ferromagnetic
ground-state with fully polarized
pseudo-spin.\cite{Kuklov,Leggett1/2} We emphasize that this is not the 
case for the double-well many-body ground-state (\ref{PsiNn}) that describes
the spin-orbit interacting BEC.  
All the arguments used for proving the ferromagnetic nature of the ground-state
for a two-component system\cite{Leggett1/2} are now irrelevant,  
as the real-space and spin components of the wave-function cannot be
factorized due to the spin-orbit coupling. The non-interacting
ground-state (\ref{PsiNn}) has an $(N+1)$-fold degeneracy. We show
bellow that this large degeneracy is partially lifted by
interactions and reduced to a two-fold degeneracy. We assume a
density-density interaction $ \hat{{\cal H}}_{\rm int} = {1 \over
2} \int d^3{\bf r} d^3{\bf r}' \hat{n} ({\bf r}) V_{\rm int}({\bf
r} - {\bf r}') \hat{n} ({\bf r}')$, where $\hat{n} ({\bf r}) = \sum_{\mu}
\hat{\psi}_\mu^\dagger ({\bf r}) \hat{\psi}_\mu ({\bf r})$ and $\hat{\psi}_\mu
({\bf r})$ is the field operator, which is initially defined in
terms of the creation/annihilation operators for the original
hyperfine states. First, we perform the position-dependent
rotation $R_{\mu \alpha}({\bf r})$ to obtain the effective
interaction term, which has the standard form
\begin{equation}
\hat{{\cal H}}_{\rm int} = {1 \over 2V} \sum\limits_{{\bf p},{\bf
p}',{\bf q}} V_{\rm int}({\bf q}) \hat{b}_{\alpha {\bf p}}^\dagger
\hat{b}_{\alpha {\bf p}+{\bf q}} \hat{b}_{\beta {\bf p}'}^\dagger
\hat{b}_{\beta {\bf p}'-{\bf q}},      \label{H_int}
\end{equation}
where $\hat{b}_{\alpha {\bf p}}^\dagger$ is the creation operator
for a state with momentum ${\bf p}$ and pseudo-spin $\alpha$ in
the dark state subspace. We need to perform a second
momentum-dependent transformation defined by  (\ref{u.up}) and
(\ref{u.down}), which introduces new bosonic operators labelled by
the band index $\lambda=\pm 1$~(\ref{E}): $\hat{B}_{\lambda {\bf
p}} = {U}^\dagger_{\lambda \alpha}({\bf p}) \hat{b}_{\alpha {\bf
p}}$ and $\hat{B}_{\lambda {\bf p}}^\dagger = \hat{b}_{\alpha {\bf
p}}^\dagger {U}_{ \alpha \lambda}({\bf p})$, where 
${U}_{ \alpha \lambda}({\bf p})={U}_{ \alpha \lambda}(\chi_{\bf p})$ and the summation  over the spin index $\alpha$ is implied.
In the limit of
relatively weak interactions, $V_{\rm int} \ll mv^2/2$ (we
emphasize that the spin-orbit coupling strength can be tuned to be
arbitrarily strong by adjusting the properties of laser fields),
the upper band with $\lambda = +1$ is irrelevant for the
low-energy physics. Thus, it is convenient to  express the
Hamiltonian in terms of left/right-moving operators, defined as
$\hat{B}_{L/R~ {\bf q}} = \hat{B}_{-1~ \mp({\bf q}+m{\bf v})}$.
Correspondingly, we have $U_{L/R ~\alpha}({\bf q}) = U_{-1
~\alpha}(\mp({\bf q}+m{\bf v}))$. This leads to the following
interaction  Hamiltonian
\begin{eqnarray}
\label{nn} \hat{{\cal H}}_{\rm int} = {1 \over 2V}\!\!
\sum\limits_{{\bf p},{\bf p}',{\bf
q}}\sum_{\{\sigma_i\}}^{~~~~~\prime}
 &&\!\!\!\!\!\!\!\! V_{\rm int}({\bf q}_{\sigma}) \hat{B}_{\sigma_1 {\bf p}}^\dagger
\hat{B}_{\sigma_2 {\bf p}+{\bf q}} \hat{B}_{\sigma_3 {\bf
p}'}^\dagger \hat{B}_{\sigma_4 {\bf p}'-{\bf q}}~~\nonumber\\ 
&& \!\!\!\!\!\!\!\!\!\!\!\!\!\!\!\!\!\!\!\!\!\!\!\!\!\!\!\!\!\!\!\!\!\!\!\!\!\!\!\!\!\!\!\!\!\!\!\!\!\!\!\!\!\times
{U}^\dagger_{\sigma_1 \alpha}({\bf p}) {U}_{\alpha\sigma_2}({\bf
p} + {\bf q}) {U}^\dagger_{\sigma_3 \alpha'}({\bf p}^\prime)
{U}_{\alpha'\sigma_4}({\bf p}^\prime - {\bf q}).
\end{eqnarray}
where the prime sign in the sum over the left and right indices
$\sigma_i = {\rm L/R}= \mp $ is restricted by the condition
$\sigma_1+\sigma_3=\sigma_2+\sigma_4$, i.e., the numbers of left-
and right-movers are conserved, and ${\bf q}_{\sigma} = {\bf q} -
 (\sigma_1-\sigma_2)m v {\bf e}_x$. We stress that equation
 (\ref{nn}) is valid in the limit of weak interactions (relative to
 the spin-orbit coupling) and low temperatures, when only
 single particle states with momenta in the vicinity of the two
 minima are occupied.

\subsection{Generalized Bogoliubov transformation} \label{IIIa}

Next, we introduce the projection operators $\hat{\cal P}_{N, n} =
\hat{\cal P}_{N, n}^2$ that select the subspace characterized by
$n$ left-moving and $(N-n)$ right-moving quasiparticles. The
Hamiltonian can be expressed as $\hat{\cal H} = \sum_{n=0}^N
\hat{\cal P}_{N, n} \hat{\cal H} \hat{\cal P}_{N, n} =
\sum_{n=0}^N \hat{\cal H}_n$. An important observation is that the
Hamiltonian containing the interaction term (\ref{nn}) preserves
the number of left- and right-movers and thus we can consider
different ``sectors,'' $\hat{\cal H}_n$, independently. Our goal
is to diagonalize each term $\hat{\cal H}_n$ using a mean-field
scheme and reduce the many-body Hamiltonian to the form
\begin{equation}
\hat{\cal H} = \sum_{n=0}^N \hat{\cal P}_{N,n} \left[{\cal E}_0(n)
+\sum_{{\bf q}, \sigma}\Omega_{\sigma}(n, {\bf
q})\hat\beta_{n,\sigma,{\bf q}}^\dagger\hat\beta_{n,\sigma,{\bf
q}}\right]\hat{\cal P}_{N,n}, \label{mfHam}
\end{equation}
where ${\cal E}_0(n)$ is the contribution of the $(n, N-n)$ sector
to the condensate energy, while $\Omega_{\sigma}(n, {\bf q})$
represents the spectrum of quasi-particle excitations. To obtain
the mean-field result, we use a Bogoliubov-type approximation in
which the operators corresponding to ${\bf q} = {\bf 0}$ are
replaced within each sector $(n, N-n)$ by $c$-numbers, $
\hat{B}_{L~0} \rightarrow \sqrt{n_0} ~e^{i \phi/2}$ and
$\hat{B}_{R~0} \rightarrow \sqrt{N_0-n_0} ~e^{-i \phi/2}$. Next,
we notice that at low temperatures, the momenta of uncondensed
bosons are $q \ll mv$. Thus, we can expand the products of
$U$-vectors in (\ref{nn}) in terms of the deviations ${\bf q}$
from the minima of the energy bands 
{\small 
\begin{eqnarray}
\vec{U}_{\rm L}^{\dagger}({\bf q}_1) \vec{U}_{\rm L}({\bf q}_2)
&=& \vec{U}_{\rm R}^{\dagger}({\bf q}_1) \vec{U}_{\rm R}({\bf
q}_2) \approx 1 - \frac{\Delta^2}{8}\frac{(q_{1y}-q_{2y})^2}{(m
v)^2},
\nonumber \\
\vec{U}_{\rm R}^{\dagger}({\bf q}_1) \vec{U}_{\rm L}({\bf q}_2)
&=& \vec{U}_{\rm L}^{\dagger}({\bf q}_1) \vec{U}_{\rm R}({\bf
q}_2) \approx \frac{\Delta}{2} \frac{q_{1y} + q_{2y}}{m v}, \label{UU_approx}
\end{eqnarray}}

\noindent with $\Delta = v^{\prime}/v < 1$ and  corrections of 
order ${\cal O}(q_{1,2}^3)$
and ${\cal O}(q_{1,2}^2)$, respectively.
Consequently, contributions to
the mean-field Hamiltonian can be expanded in the small parameter
$x_{\bf q} = \Delta^2q_y^2/(m v)^2$. In the zero-order
approximation, i.e., neglecting contributions of order $x_{\bf q}$
and higher, the mean-field Hamiltonian for the $(n, N-n)$ sector is 
{\small
\begin{eqnarray}
\hat{{\cal H}}_n^{(0)} &=& {N \over 2V} \sum\limits_{\bf q} V_{\rm
int}({\bf q}) \left[ \hat{\vec{B}}^\dagger_{\bf q}\left(
\begin{array}{cc}
s({\bf q}) + 1 + \delta & \sqrt{1-\delta^2} e^{-i \phi}  \\
\sqrt{1-\delta^2} e^{i \phi} &  s({\bf q}) + 1 -\delta  \\
 \end{array} \right)\hat{\vec{B}}_{\bf q} \right.  \nonumber \\
 &+& \left. \hat{\vec{B}}^{\rm T}_{\bf q} \left(
\begin{array}{cc}
(1+\delta) e^{i\phi} & \sqrt{1-\delta^2}  \\
\sqrt{1-\delta^2} &  (1-\delta) e^{-i\phi}  \\
 \end{array} \right) \hat{\vec{B}}_{-\bf q} + {\rm h.c.}\right],
 \label{mf}
\end{eqnarray}}

\noindent where  $\delta = 2n/N-1$, $\hat{\vec{B}}^T_{\bf q} = \left(
\hat{B}_{L {\bf q}}, \hat{B}_{R {\bf q}} \right)$ is the
annihilation operator in a spinor notation, $s({\bf q}) = 2 \delta
E({\bf q})/\left[ n_0 V_{\rm int}({\bf q})\right]$, and $\delta
E({\bf q})$ is the anisotropic spectrum (\ref{delE}) near the minima. We now
introduce new bosonic operators $ \hat{B}_{-,{\bf q}} =
\sqrt{1-n/N} \hat{B}_{L,{\bf q}} e^{-i \phi /2} - \sqrt{n/N}
\hat{B}_{R,{\bf q}} e^{i \phi /2}$ and $ \hat{B}_{+,{\bf q}} =
\sqrt{n/N} \hat{B}_{L,{\bf q}} e^{-i \phi /2} + \sqrt{1-n/N}
\hat{B}_{R,{\bf q}} e^{i \phi /2}$. The Hamiltonian  becomes
diagonal for the $\hat{B}_-$-particles, which have the ``free''
spectrum $\delta E({\bf q})$, and has the standard Bogoliubov
form~\cite{AGD} for the $\hat{B}_+$-particles.
 Introducing the new operators $\hat{\beta}_{-, {\bf q}} \equiv
\hat{B}_{-, {\bf q}}$ and $\hat{\beta}_{+, {\bf q}} \equiv
(1-A_{\bf q}^2)^{-1/2} \left( \hat{B}_{+, {\bf q}} + A_{\bf q}
\hat{B}_{+, -{\bf q}}^\dagger \right)$, with $A_{\bf q} = -s({\bf
q}) - 1 +\sqrt{\left[ s({\bf q}) + 1 \right]^2 - 1}$, we get
{\small
\begin{eqnarray}
\label{B+-}  \hat{{\cal H}}_n^{(0)} = {\cal E}_0^{(0)} + \sum\limits_{\bf q}
\Biggl\{ \Omega_-({\bf q}) \hat{\beta}_{-, {\bf q}}^\dagger
\hat{\beta}_{-, {\bf q}}   + \Omega_+({\bf q}) \hat{\beta}_{+,
{\bf q}}^\dagger \hat{\beta}_{+, {\bf q}} \Biggr\},
\end{eqnarray}}

\noindent where ${\cal E}_0^{(0)}$ is the condensate energy~\cite{AGD} in
the zero-order approximation, $ \Omega_-({\bf q}) = \left\{ q_x^2
+ q_z^2 + q_y^2 \left[ 1 - (v'/v)^2 \right] \right\}/(2m)$ is the
anisotropic free particle quadratic spectrum and $ \Omega_+({\bf
q}) = \sqrt{\left[  \Omega_-({\bf q}) + {n V_{\rm int}({\bf q})}
\right]^2 - n^2 V_{\rm int}^2({\bf q}) }$ is  an anisotropic sound
similar to the conventional Bogoliubov phonon mode in a BEC. At
this level of approximation the condensate energy is n-independent
(i.e., it is the same for any particular 
sector characterized by n left movers and ($N-n$) right movers)
and, consequently,  the degeneracy of the non-interacting ground
state (\ref{PsiNn}) is preserved. In the first order
approximation, the mean-field Hamiltonian (\ref{mf}) acquires
sector-dependent corrections of order $x_{\bf q}\ll 1$. Following
the above recipe, we introduce a set of new operators
$\hat{B}_{\pm,{\bf q}}$ that diagonalize the
$\hat{\vec{B}}^\dagger_{\bf q}\hat{\vec{B}}_{\bf q}$ term in the
Hamiltonian~(\ref{mf}) but not the other terms. Next, we
diagonalize  the full Hamiltonian [up to terms of order ${\cal
O}(x_{\bf q}^2)$] via a generalized Bogoliubov-type transformation
\begin{eqnarray}
\hat{\beta}_{-, ~{\bf q}} &=& \hat{B}_{-, {\bf q}} + x_{\bf q}D_{\bf q} \hat{B}_{-, -{\bf q}}^\dagger  \label{bet_m}\\
&& ~~~~~~~~~~~
+ x_{\bf q} F_{1\bf q}\hat{B}_{+, {\bf q}} + x_{\bf q} F_{2\bf q} \hat{B}_{+, -{\bf
q}}^\dagger \nonumber \\
\hat{\beta}_{+, ~{\bf q}} &=& (1-A_{\bf q}^2)^{-1/2} \left(
\hat{B}_{+, {\bf q}} + A_{\bf q} \hat{B}_{+, -{\bf q}}^\dagger \right . \label{bet_p} \\ 
&& ~~~~~~~~~~~
+  \left.x_{\bf q} C_{1\bf q} \hat{B}_{-, {\bf q}} + x_{\bf q} C_{2\bf q}
\hat{B}_{-, -{\bf q}}^\dagger\right). \nonumber
\end{eqnarray}
In the equations (\ref{bet_m}) and (\ref{bet_p}) we already anticipated
that some of the terms are corrections of order $x_{\bf q}$. The
coefficients are determined by requiring that the
$\beta$-operators obey standard commutation relations [to order
${\cal O}(x_{\bf q})$] and that the off-diagonal contributions to
the Hamiltonian vanish. Assuming for simplicity  that we have a
point-like interaction, i.e., $V_{\rm int}({\bf q})=V_{\rm int}$
is momentum-independent for momenta in a range that is relevant
for the problem, the  groundstate energy in the (n, N-n) sector
is
{\small
\begin{eqnarray}
{\cal E}_0(n) &=& \frac{V_{\rm int}N^2}{2V} + \frac{V_{\rm
int}N}{2V}\sum_{{\bf q}\neq 0} \frac{1}{1-A_{\bf q}^2}\left\{
[2+s({\bf q})]A_{\bf q}^2 + 2 A_{\bf q} \right. \nonumber \\
&& \!\!\!\!\!\!\!\!\!\!\!\!\!\!\!\!\!\!\!\! - \left.\frac{x_{\bf
q}}{8}\left[A_{\bf q}^2(\cos(4\xi)+3) - A_{\bf
q}(\cos(4\xi)-5)\right]\right\} + {\cal O}(x_{\bf q}^2),
\label{E_0}
\end{eqnarray}}

\noindent where $\cos^2(\xi) = n/N$. The relevant coefficient of the
generalized Bogoliubov transformation (\ref{bet_m}-\ref{bet_p})
has the form
{\small
\begin{eqnarray}
A_{\bf q} &=& -1 -\frac{s}{2} + \frac{1}{2}\sqrt{s(4+s)} -
\frac{x_{\bf q}}{32\sqrt{s(4+s)}}
\left(2+s\sqrt{s(4+s)}\right) \nonumber \\
&& \times\left[-4-5s+(4+s)\cos(4\xi)\right] + {\cal O}(x_{\bf q}^2).    \label{A_q}
\end{eqnarray}}

\noindent Explicitly evaluating (\ref{E_0}) with $A_{\bf q}$ given by Eq.
(\ref{A_q}) shows that, at this level of approximation, the energy
of the condensate becomes sector-dependent, ${\cal E}_0(n) \approx
{\cal E}_0^{(0)} + {\cal E}_0^{(1)}(n)$,  and is minimal for $n=0$
and $n=N$. Thus, the density-density interaction reduces the large
$(N+1)$-fold degeneracy of the ground state to a two-fold
degeneracy. Consequently, in the limit of vanishing interactions
$V_{\rm int} \to +0$, the most general expression for the
many-body wave-function is
{\small
\begin{equation}
||\Psi_N\rangle = \frac{1}{\sqrt{N!}}\left[\sqrt{w_{\rm
L}}e^{i\phi_{\rm L}}\left(\hat{B}_{\rm L}^{\dagger}\right)^N
+\sqrt{w_{\rm R}}e^{i\phi_{\rm R}}\left(\hat{B}_{\rm
R}^{\dagger}\right)^N\right]||{\rm vac}\rangle, \label{GSN}
\end{equation}}

\noindent where $w_{\rm L/R}$ represents the fraction of the left/right
movers and $\phi_{\rm L/R}$ are arbitrary phases. Notice that
Eq.~(\ref{GSN}) describes a fragmented or entangled BEC, unless
$w_{\rm L}w_{\rm R} = 0$. I.e., the many-body state (\ref{GSN}) does not
correspond to the condensation into one single-particle state.
We reiterate that the left- and right-movers in
the condensate have non-zero momentum, but zero velocity and do
not actually move while the laser fields responsible for the
spin-orbit coupling are present. We also note that equation (\ref{GSN})
describes a so-called NOON state,\cite{NOON,NOON1} which is quantum 
correlated state with properties that can be exploited in applications 
such as quantum sensing and quantum metrology. This suggests that the 
possibility of using spin-orbit coupled condensates as qubits deserves 
to be further investigated.

\subsection{Gross-Pitaevskii equations} \label{IIIb}

Let us consider the density-density interaction potential as a contact
pseudo-potential, 
$V_{\rm int}({\bf r} - {\bf r}')=V_{\rm int}\delta({\bf r} - {\bf r}')$, 
where $V_{\rm int}=\frac{4\pi\hbar^2}{m}a$ and  $a$ is the inter-atomic scattering length. The full many body Hamiltonian can be written as 
\begin{eqnarray}
\hat{\cal H}&=&\sum_{\mu, \nu}\int {d}^3{r} ~\hat{\psi}^\dagger_{\mu} ({\bf r}) h_{\mu\nu} \hat{\psi}^{ }_{\nu} ({\bf r}) \label{fullH}\\
&& + \frac{V_{\rm int}}{2} \sum_{\mu, \nu}\int{d}^3{r} ~\hat{\psi}^\dagger_{\mu} ({\bf r}) \hat{\psi}^\dagger_{\nu} ({\bf r}) \hat{\psi}_{\nu} ({\bf r}) \hat{\psi}_{\mu}({\bf r}),  \nonumber 
\end{eqnarray}
in terms of field operators  $\hat{\psi}_{\mu}({\bf r})$  for the original hyperfine states,  $\mu\in\{0, 1, 2, 3\}$. In Eq. (\ref{fullH}) we used  the notation ${h}_{\mu\nu}=\left\{ \frac{{\bf p}^2}{2m} + V_{\rm trap} + H_{a-l}\right\}_{\mu\nu}$ for the single particle Hamiltonian in the presence of a trap potential $V_{\rm trap}$, in addition to the spatially varying laser fields that interact with the atom, $ H_{a-l}$. In the adiabatic approximation, after projecting onto the dark state subspace, the first term in Eq. (\ref{fullH}) becomes $\sum\limits_{{\bf p};\alpha,\beta} \hat{b}_{\alpha {\bf p}}^{\dagger} \left\{ \left[{\bf p}^2/2m+ V_{\rm trap}\right]\check{1} - v p_x \check{\sigma}_2 - v^{\prime} p_y\check{\sigma}_3 \right\}_{\alpha \beta}  \hat{b}_{\beta {\bf p}}$, where $\hat{b}_{\alpha {\bf p}}^{\dagger}$ and $\hat{b}_{\alpha {\bf p}}$ are the creation and annihilation operators for bosons with pseudo-spin $\alpha ={\uparrow,\downarrow}$. The interaction term is given by equation (\ref{H_int}). Before writing down the Gross-Pitaevskii equations, let us summarize the three different representations used for describing the system of bosons interacting with the spatially modulated laser fields.

i) {\it Hyperfine states representation}: This is the most straightforward way to describe the motion of the bosons and their interaction with the trap potential ($V_{\rm trap}$) and the laser fields ($H_{a-l}$), as well as the density-density  interaction (second term in Eq. (\ref{fullH})). The field operator that creates a particle in the hyperfine state $\mu\in\{0, 1, 2, 3\}$ at point ${\bf r}$ is $\hat{\psi}_{\mu}^{\dagger}({\bf r})$, while the creation of a free-moving particle with momentum ${\bf p}$ is described by $\hat{c}_{\mu{\bf p}}^{\dagger} = \int d^3r~ e^{i {\bf p r}}\hat{\psi}_{\mu}^{\dagger}({\bf r})$.  By performing the position-dependent rotation $R_{\mu\alpha}$ which diagonalizes the atom-laser Hamiltonian {\it and} projecting onto the dark states subspace we switch to the pseudo-spin representation. 

ii) {\it Pseudo-spin representation (dark states representation)}: This is the natural framework for describing the low-energy physics of the atomic system interacting with the laser field. The creation operator for free-moving particles with spin $\alpha\in\{\uparrow, \downarrow\}$ and momentum {\bf p} is $\hat{b}_{\alpha{\bf p}}^{\dagger}$. We can define the corresponding field operator as $\hat{\widetilde{\psi}}_{\alpha}^{\dagger}({\bf r}) = \sum_{\bf p} e^{-i{\bf p r}} \hat{b}_{\alpha{\bf p}}^{\dagger}$.  Note that the field operators in the hyperfine and pseudo-spin representations are related via the position-dependent rotation, $\hat{\psi}_{\mu}^{\dagger}({\bf r}) = \sum_{\alpha} R_{\mu \alpha}({\bf r})   \hat{\widetilde{\psi}}_{\alpha}^{\dagger}({\bf r})$.  Diagonalizing the single-particle spin-orbit coupled Hamiltonian,  $H =  \left[{\bf p}^2/2m+ V_{\rm trap}\right]\check{1} - v p_x \check{\sigma}_2 - v^{\prime} p_y\check{\sigma}_3$, generates  a set of eigenstates described by the spinor eigenfunctions $\vec{\phi}_{\sigma n}({\bf r})$. The quantum number $\sigma=\pm$ can be viewed as labeling right (left) moving states.

iii) {\it Right/left moving states representation}: This is the representation corresponding to the eigenstates of the spin-orbit coupled single particle Hamiltonian. In Section \ref{I} we have shown that in the absence of a trap potential the single particle spectrum for the generic case $v \neq v^{\prime}$ is characterized by two minima at non-zero momenta. Here we show explicitly that the double-degeneracy of the single-particle states is a general property of the spin-orbit interacting Hamiltonian, protected by a Kramers-like symmetry. Let us use the following parametrization for the eigenfunctions:
\begin{equation}
\vec{\phi}_{\sigma n}({\bf r}) = e^{i\sigma m v x}\left( 
\begin{array}{c}
u_{\sigma n}^{\uparrow}({\bf r}) \\ i \sigma u_{\sigma n}^{\downarrow}({\bf r}) 
\end{array}
\right), \label{phi_sn}  
\end{equation}
where $\sigma=\pm$ and n is a set of quantum numbers. The components $u_{\sigma n}^{\alpha}({\bf r})$ are the solutions of the following eigenproblem
\begin{eqnarray}
\left(
\begin{array}{cc}
h_0 - v^{\prime}p_y & i v p_x \\
- i v p_x & h_0 + v^{\prime}p_y
\end{array}
\right) && \!\!\!\!\!\!\!\left(
\begin{array}{c}
u_{\sigma n}^{\uparrow}({\bf r}) \\ i \sigma u_{\sigma n}^{\downarrow}({\bf r}) 
\end{array}
\right) e^{i\sigma m v x} \nonumber \\
&&\!\!\!\!\!\!\!\!\!\!\!\!\!\!\!\!\!\!\!\!\!\!\!\!\!\!\!\! = E_{\sigma n} \left( 
\begin{array}{c}
u_{\sigma n}^{\uparrow}({\bf r}) \\ i \sigma u_{\sigma n}^{\downarrow}({\bf r}) 
\end{array}
\right) e^{i\sigma m v x},                        \label{eigensyst}  
\end{eqnarray}
where $h_0 = p^2/2m + V_{\rm trap}$ is the Hamiltonian in the absence of spin-orbit interaction. More explicitly, $u_{\sigma n}^{\alpha}({\bf r})$  satisfy the following system of coupled differential equations:
\begin{eqnarray}
\left[-\frac{\nabla^2}{2m}+V_{\rm trap}({\bf r}) +iv^{\prime}\frac{\partial}{\partial y} - E -\frac{m v^2}{2}\right] u_{\sigma}^{\uparrow}({\bf r}) && \nonumber \\
+ \left[-i \sigma v \frac{\partial}{\partial x} + mv^2\right] \left( u_{\sigma}^{\uparrow}({\bf r}) - u_{\sigma}^{\downarrow}({\bf r}) \right) &=& 0, \nonumber \\
\left[-\frac{\nabla^2}{2m}+V_{\rm trap}({\bf r}) -iv^{\prime}\frac{\partial}{\partial y} - E -\frac{m v^2}{2}\right] u_{\sigma}^{\downarrow}({\bf r}) && \label{u_up_down}\\
- \left[-i \sigma v \frac{\partial}{\partial x} + mv^2\right] \left( u_{\sigma}^{\uparrow}({\bf r}) - u_{\sigma}^{\downarrow}({\bf r}) \right) &=& 0 . \nonumber 
\end{eqnarray}
Taking the complex conjugate of (\ref{u_up_down}) with $\sigma \rightarrow -\sigma$ we obtain an identical set of equations. Consequently  we have
\begin{eqnarray}
u_{-{\sigma} n}^{\uparrow}({\bf r}) &=& \left[u_{\sigma n}^{\downarrow}({\bf r})\right]^*,     \nonumber \\
u_{-{\sigma} n}^{\downarrow}({\bf r}) &=& \left[u_{\sigma n}^{\uparrow}({\bf r})\right]^*,       \label{u_minsig}
\end{eqnarray}
and the corresponding energies are degenerate, $E_{-{\sigma} n} = E_{\sigma n} = E_n$. Because $\langle \phi_{-{\sigma}n}| \phi_{\sigma n} \rangle =0$, the two states are linearly independent. We conclude that the single-particle  eigenstates of the spin-orbit coupled Hamiltonian are (at least) double degenerate independent of the symmetries (or lack of symmetry) of the trap potential. Note that this double degeneracy is a consequence of a Kramers-like symmetry of the spin-orbit interacting Hamiltonian, which contains terms that are either quadratic in momentum, or linear in both momentum and spin. The creation operator for a left/right moving particle described by the eigenstate $\vec{\phi}_{\sigma n}$ is $\hat{B}_{\sigma n}^\dagger$. The field operators in the pseudo-spin representation can be expressed in terms of $\hat{B}_{\sigma n}$ operators  as
{\small
\begin{eqnarray}
\hat{\widetilde{\psi}}_{\uparrow}({\bf r}) &=& \sum_{n} \left[~e^{i m v x}u_{+ n}^{\uparrow}({\bf r}) \hat{B}_{+ n} + ~e^{- i m v x}u_{- n}^{\uparrow}({\bf r}) \hat{B}_{- n} \right],   \nonumber \\
 \hat{\widetilde{\psi}}_{\downarrow}({\bf r}) &=& \sum_{n} \left[i e^{i m v x}u_{+ n}^{\downarrow}({\bf r}) \hat{B}_{+ n} - i e^{- i m v x}u_{- n}^{\downarrow}({\bf r}) \hat{B}_{- n} \right],
\end{eqnarray}}

\noindent where the terms with $\sigma = +$ and $\sigma = -$ correspond to the right and left moving modes, respectively. Finally, note that in the translation invariant case, $V_{\rm trap}=0$, we introduced the eigenfunctions $\vec{\phi}_{\lambda{\bf p}}({\bf r})=  e^{i{\bf p r}} \vec{U}_{\lambda}(\chi_{\bf p})$, with ${U}_{\alpha \lambda}(\chi_{\bf p})$ given by equations (\ref{u.up}) and (\ref{u.down}), and  the corresponding creation operators, $\hat{B}_{\lambda {\bf p}}$.  We then  defined the left/right movers for the low energy band $\lambda = -1$ and small momenta $q<mv$ as $\hat{B}_{L/R~ {\bf q}} = \hat{B}_{(-1)~ \mp({\bf q}+m{\bf v})}$. Alternatively, we can directly define the eigenfunctions $\vec{\phi}_{\sigma {\bf q}}({\bf r})$ in the left/right moving representation using the parametrization  (\ref{phi_sn}), with no restriction for ${\bf q}$. The correspondence between the two representations is given by: ${\bf p}=\sigma({\bf q} + m{\bf v})$ and $\lambda = -{\rm sign}(q_x + mv)$. This generalizes our definition of the left/right moving modes to arbitrary energy. Notice however, that a left (right) ``moving'' state from the high energy band $\lambda=+1$ has in fact a positive (negative) momentum.

To write the Gross-Pitaevskii equation in the pseudo-spin representation we use the standard procedure and calculate the commutator $[ \hat{\widetilde{\psi}}_{\alpha}({\bf r}) , \hat{\cal H} ]$, where $\hat{\cal H}$ is the many-body Hamiltonian expressed in terms of pseudo-spin field operators. Using Eq. (\ref{fullH}) and the relations between representations summarized above we obtain
{\small
\begin{eqnarray}
i\frac{\partial}{\partial t}\widetilde{\psi}_{\alpha}({\bf r}, t) &=& \sum_{\beta} \left\{ \left[\frac{-\nabla^2}{2m} + V_{\rm trap}({\bf r})\right] \check{1} + i v \frac{\partial}{\partial x} \check{\sigma}_2 \right. \label{GPE} \\
&& \!\!\!\!\!\!\!\!\!\!\!\!\!\!\!\!\!\!\!\!\!\!\!\!\!\! + \left. i v^\prime \frac{\partial}{\partial y} \check{\sigma}_3 \right\}_{\alpha\beta} \widetilde{\psi}_{\beta}({\bf r}, t) + V_{\rm int} \left( |\widetilde{\psi}_{\uparrow}|^2 + |\widetilde{\psi}_{\downarrow}|^2\right)\widetilde{\psi}_{\alpha}({\bf r}, t).   \nonumber   
\end{eqnarray}}
 
\noindent Relation (\ref{GPE}), which is a system of two coupled non-linear differential equations, represents the time-dependent Gross-Pitaevskii equation for a spin-orbit coupled Bose-Einstein condensate wave-function. Similar equations can be written in the left/right moving states representation. For simplicity, we will address here only the translation invariant case $V_{\rm trap} =0$. The field operator for the left/right moving modes can be written in terms of the corresponding $\hat{B}_{\sigma {\bf q}}$ operators as
\begin{equation}
\hat{\widetilde{\widetilde{\psi}}}_{\sigma}({\bf r}) = \sum_{\alpha, {\bf q}}~\phi_{\sigma {\bf q}}^{\alpha}({\bf r}) \hat{B}_{\sigma {\bf q}}.
\end{equation}
The non-interacting part of the Hamiltonian is diagonal in terms of left/right moving operators, with eigenvalues that depend on the momentum ${\bf q}$ only. At low-energies, these eigenvalues are given by the anisotropic spectrum $\delta E({\bf q}) = (q_x^2+q_z^2)/(2m) + q_y^2/(2m_y)$ with $m_y = m\left[1-(v^\prime/v)^2\right]^{-1}$. In general, the interacting Hamiltonian is given by equation (\ref{nn}), but in the low-energy limit we neglect all corrections of order $x_{\bf q} = \Delta^2q_y^2/(mv)^2$ and higher coming from the momentum-dependent matrices $U_{\alpha\sigma}({\bf q})$.  In this limit we obtain
\begin{eqnarray}
i\frac{\partial}{\partial t}\widetilde{\widetilde{\psi}}_{\sigma}({\bf r}, t) &=& \left(\frac{(-i\partial_x - \sigma m v)^2}{2m} - \frac{\partial_y^2}{2m_y} -  \frac{\partial_z^2}{2m}\right)\widetilde{\widetilde{\psi}}_{\sigma}({\bf r}, t) \nonumber \\
&& + V_{\rm int} \left(|\widetilde{\widetilde{\psi}}_{L}|^2 + |\widetilde{\widetilde{\psi}}_{R}|^2\right)\widetilde{\widetilde{\psi}}_{\sigma}({\bf r}, t),
\end{eqnarray}
where $\partial_j = \partial/\partial x_j$, $j\in\{x, y, z\}$. The time-independent Gross-Pitaevskii equations can be obtained by looking for a stationary solution of the form $\widetilde{\widetilde{\psi}}_{\sigma}({\bf r}, t) = \widetilde{\widetilde{\psi}}_{0 \sigma}({\bf r})e^{-i\mu t}$, where $\mu$ is the chemical potential which determined by the condition $N = \int d^3r\left( |\widetilde{\widetilde{\psi}}_{L}|^2 + |\widetilde{\widetilde{\psi}}_{R}|^2 \right)$, with $N$ being the total number of bosons. We note that by linearizing $\widetilde{\widetilde{\psi}}_{\sigma}({\bf r}, t)$ with respect to the deviations from the stationary solution we obtain an excitation spectrum consisting in two modes, $\Omega_{\pm}({\bf q})$, identical with those found using the generalized Bogoliubov treatment.

\section{Experimental signature of spin-orbit coupled BEC: measuring a SOBEC qubit} \label{IV}

A straightforward way to  detect experimentally the new type of
BEC would be to probe the momentum distribution of  the density of
the particles via time-of-flight expansion. After removing the
trap {\em and} the laser fields, the boson gas represents a system
of free particles, each characterized by a certain momentum and a
hyperfine state index. In a TOF experiment one determines the
momentum distribution by measuring the particle density at various
times after the release of the boson cloud. The operator
associated with a density measurement is $\hat{\rho}({\bf r}) =
\sum_{\mu} \hat{\psi}_{\mu}^{\dagger}({\bf
r})\hat{\psi}_{\mu}({\bf r})$, where
$\hat{\psi}_{\mu}^{\dagger}({\bf r})$ is the creation operator
for a particle in the  hyperfine state $\mu$ positioned at
point ${\bf r}$. Determining the density profile involves a
simultaneous measurement of $\hat{\rho}({\bf r})$ for all the
values of ${\bf r}\in{\cal V}$ corresponding to a ceratin region
in space where the boson cloud is located. To insure formal
simplicity, we consider a coarse-grained space, i.e., we treat
${\bf r}$ as a discrete variable. This is simply a technical trick 
and does not influence the final result. Our goal is to find the most 
likely spatial distributions of the particles at a given moment $t$ after 
the release of the atoms.  In the limit of large particle numbers, the 
actual  measured density profiles will involve only small fluctuations 
away from these ``most likely'' distributions.

For a system of N bosons, the
result of the measurement is a set of eigenvalues
$\{\sum_{\mu}n_{{\bf r} \mu}\}_{({\bf r}\in{\cal V})}$ that
label an eigenvector of the density operator
\begin{equation}
||\Phi_{\{n_{{\bf r}\mu}\}} \rangle = \prod_{\mu, {\bf r}\in
{\cal V}}\frac{1}{\sqrt{(n_{{\bf r}\mu})!}}\left[\hat{\psi}_{\mu}^{\dagger}({\bf
r})\right]^{n_{{\bf r}\mu}}||{\rm vac}\rangle,
\label{dens_eigenstate}
\end{equation}
where the occupation numbers satisfy the constraint $\sum_{\mu,
{\bf r}}n_{{\bf r}\mu} = N$, and the factors $1/\sqrt{(n_{{\bf r}\mu})!}$ insure the normalization to unity. Note that
$n_{{\bf r}\mu}$ is an integer representing the number of
particles located in a certain ``cell'' ${\bf r}$ of the
coarse-grained space. At time t after the release, the many-body
state of N bosons that were initially in a BEC groundstate
described by Eq. (\ref{GSN}) is
\begin{eqnarray}
||\widetilde{\Psi}_N(t)\rangle &=& {\cal N} \sum_{\sigma}
\sqrt{w_{\sigma}}~e^{i\phi_{\sigma}} \label{PsiN_t} \\
&& \!\!\!\!\!\!\!\!\!\!\!\!\!\! \sum_{\{n_{{\bf
r}\mu}\}_{\cal V}}\left\{\prod_{\mu, {\bf r}\in{\cal V}}
\frac{1}{(n_{{\bf r}\mu})!}\left[Q_{\mu}^{\sigma}({\bf r}, t)
\hat{\psi}_{\mu}^{\dagger}({\bf r})\right]^{n_{{\bf r}\mu}}
||{\rm vac}\rangle\right\},  \nonumber 
\end{eqnarray}
where ${\cal N}$ is a normalization factor, $\sigma$ labels the
left ($\sigma=L\equiv -1$) and right ($\sigma=R\equiv +1$) modes
and $||\widetilde{\Psi}_N(0)\rangle = ||\Psi_N\rangle$. The
coefficients $Q_{\mu}^{\sigma}$ are normalized so that
$\sum_{{\bf r},\mu}|Q_{\mu}^{\sigma}({\bf r}, t)|^2 = 1$.
The second summation in (\ref{PsiN_t}) is over all possible
spatial distributions of N particles and, in the continuous limit,
it becomes a path integral. Equation (\ref{PsiN_t}) represent the
expansion of the many-body wave-function in terms eigenstates
(\ref{dens_eigenstate}) of the density operator. The probability
${\cal P}[\{n_{{\bf r}\mu}\}]$ of measuring a certain density
profile $n_{{\bf r}\mu}$ is determined by the coefficient of
the corresponding term. If we focus, for simplicity, on the case
when there are only left (right) movers in (\ref{GSN}), this
probability is proportional to $\prod_{{\bf
r},\mu}|Q_{\mu}^{\sigma}({\bf r}, t)|^{2 n_{{\bf
r}\mu}}/(n_{{\bf r}\mu})!$, with $\sigma = L (R)$. The
probability ${\cal P}[\{n_{{\bf r}\mu}\}]$ has a maximum for
$n_{{\bf r}\mu}^0 = N |Q_{\mu}^{\sigma}({\bf r}, t)|^2$
corresponding, in the continuous limit, to a stationary point of
the path integral in equation (\ref{PsiN_t}). For large particle
number, ${\cal P}[\{n_{{\bf r}\mu}\}]$ becomes sharply peaked
at $n_{{\bf r}\mu}^0$ and the actually measured density
profiles will exhibit only relatively small deviations from the
stationary profile. Therefore, at time t after release, the
density of the boson cloud is
\begin{equation}
\rho({\bf r}, t) = N\sum_{\mu}~|Q_{\mu}^{\sigma}({\bf r},
t)|^2.   \label{rho_rt}
\end{equation}
If both $w_R$ and $w_L$ are non-zero, the result of a measurement
will be either a ``right moving'' density profile [$\sigma = R$
in (\ref{rho_rt})] with a probability $w_R$, or a ``left moving''
profile [$\sigma = L$ in (\ref{rho_rt})] with a probability
$w_L$, assuming that the two profiles are spatially well
separated. We are not addressing here the interesting effects of
the interference between left and right moving condensates.
These effects
are negligible if the left and right moving density profiles are
spatially separated, but become important otherwise, e.g. at small
times after the release.

Next we determine explicitly the coefficients $Q_{\mu}^{\sigma}({\bf r}, t)$ for the exactly solvable model of bosons with ``Ising-type'' spin-orbit coupling, $v \ne v^\prime=0$, placed in a harmonic trap, $V_{\rm trap}({\bf r}) = m\omega^2 r^2/2$.\cite{SZG} In this case, the operators
$\hat{B}_{\sigma}^{\dagger}$ from Eq. (\ref{GSN}) are creation
operators for the single particle ground states
\begin{equation}
\vec{\phi}_{\sigma 0}({\bf r}) = \varphi_0({\bf r}) e^{i\sigma m
v x} \frac{1}{\sqrt{2}}\left( \begin{array}{c} 1 \\ i\sigma
\end{array}\right), \label{psi_trap}
\end{equation}
where $\varphi_0({\bf r})$ represents the groundstate wavefunction
of the harmonic oscillator. The spinor (\ref{psi_trap}) is written
in the dark state basis. Performing the position-dependent
rotation $R_{\mu\alpha}$ [see Eq. (\ref{darkSts})], we can
express the operators $\hat{B}_{\sigma}^{\dagger}$ in terms of
creation operators for particles in a certain hyperfine state
located at point ${\bf r}$, $\hat{\psi}_{\mu}^{\dagger}({\bf
r})$, or their Fourier components corresponding to free moving
particles, $\hat{c}_{\mu{\bf k}}^{\dagger} = \sum_{\bf r}
e^{i{\bf k r}}\hat{\psi}_{\mu}^{\dagger}({\bf r})$. The time
evolution after the release can be easily described in terms of
time evolution for the $\hat{c}_{\mu{\bf k}}^{\dagger}$
operators, $\hat{c}_{\mu{\bf k}}^{\dagger}(t) =
\exp(-i\epsilon_{\bf k} t)~\hat{c}_{\mu{\bf k}}^{\dagger}$,
where $\epsilon_{\bf k} = k^2/(2m)$ is the free particle spectrum.
Consequently, the many-body state $||\widetilde{\Psi}_N(t)\rangle$
can be obtained by making in Eq. (\ref{GSN}) the substitution
$\hat{B}_{\sigma}^{\dagger} \rightarrow \sum_{{\bf r}, \mu}
Q_{\mu}^{\sigma}({\bf r}, t)
\hat{\psi}_{\mu}^{\dagger}({\bf r})$ with
\begin{equation}
Q_{\mu}^{\sigma}({\bf r}, t) = \sum_{\alpha, {\bf k}, {\bf
r}^{\prime}} \left[\vec{\phi}_{\sigma 0}\right]_{\alpha}({\bf
r}^{\prime}) R_{\mu\alpha}^*({\bf r}^{\prime})e^{i{\bf k}({\bf r}
-{\bf r}^{\prime})} e^{-i\epsilon_{\bf k} t}.  \label{Q_rt}
\end{equation}
Finally, introducing this expression of $Q_{\mu}^{\sigma}$ in
equation (\ref{rho_rt}) we obtain for the measured density profile
the expression
{\small
\begin{eqnarray}
\rho({\bf r}, t) &=& N
\frac{\Gamma^3}{\left[2\pi\left(1+\tau^2\right)\right]^{\frac{3}{2}}}
~e^{-\frac{\Gamma^2(y^2+z^2)}{1+\tau^2}}\left[\sin^2\theta
~e^{-\frac{\Gamma^2}{1+\tau^2}\left(x - \frac{\lambda m v
t}{m}\right)^2} \right.\nonumber \\
&&  \!\!\!\!\!\!\!\!+ \left. \frac{(1-\lambda\cos\theta)^2}{2}
~e^{-\frac{\Gamma^2}{1+\tau^2}\left(x - \frac{(\lambda m v +m v_a)
t}{m}\right)^2} \right.  \nonumber \\
&& \!\!\!\!\!\!\!\!+ \left. \frac{(1+\lambda\cos\theta)^2}{2}
~e^{-\frac{\Gamma^2}{1+\tau^2}\left(x - \frac{(\lambda m v -m v_a)
t}{m}\right)^2} \right],  \label{TOF_dens}
\end{eqnarray}} 

\noindent where $\Gamma = \sqrt{m \omega}$ is the inverse characteristic
length of the trap potential,  $\tau = \omega t$ is time in units
of $\omega^{-1}$,  $\theta\in[0\pi/2]$ and $v_a$ are tunable
parameters characterizing the laser field, and $v=v_a\cos\theta$
[see the paragraph containing Eq. (\ref{darkSts})].
In equation (\ref{TOF_dens}) the density was normalized so that
$\int d^3r~\rho({\bf r}, t) = N$.
\begin{figure}[tbp]
\begin{center}
\includegraphics[width=0.43\textwidth]{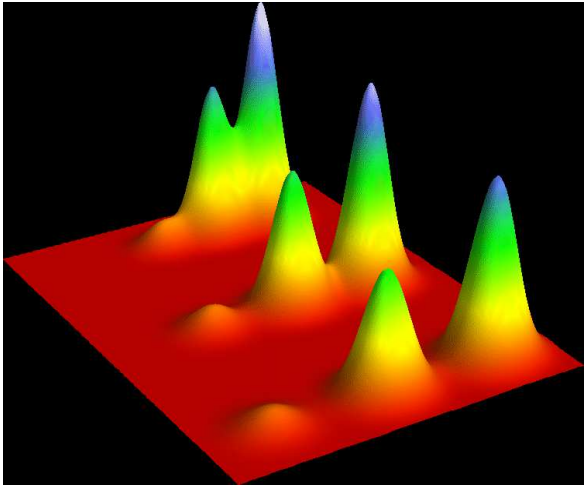}
\end{center}
\caption{(color online): Density of particles at three different
moments,  $t_1 = 0.4 \omega^{-1}$, $t_2 = 0.6 \omega^{-1}$, and
$t_3 = 0.8 \omega^{-1}$, after both the trap and the laser fields
are removed at $t=0$. For clarity, the density distributions are
shifted along the y-axis. This time-of-flight expansion
corresponds to a many-body ground state (\ref{GSN}) and is
obtained using the single-particle eigenfunctions  for the
exactly-solvable model of trapped bosons with Ising-type
spin-orbit coupling ($v \ne v'=0$)~\cite{SZG} with $v_a = 6
(\omega / m)^{1/2}$. This ``left moving'' density distribution is
measured with a probability $w_L$, while there is a $w_R$
probability to observe a ``right moving'' distribution which
corresponds to a $x\rightarrow -x$ reflection (see also Fig.
\ref{Fig3}). Notice the characteristic three-peak structure. To
resolve the BEC peaks, the spin-orbit coupling energy scale should
be larger than the trap level spacing, i.e., $mv^2 \gg \omega$. In
the opposite limit the phenomenon of real-space BEC separation is
smeared out by finite-size effects~(\ref{GSN}).} \label{Fig2}
\end{figure}
The density profile for a "left moving"  density distribution
($\sigma = -1$) is shown in Fig. \ref{Fig2} for three different
times after the release of the boson cloud. The parameters of the
calculation are $\theta = \pi/3$ and $v_a = 6\sqrt{\omega/m}$.
Notice the three-peak structure of the density, corresponding to
the three exponential terms in equation (\ref{TOF_dens}). The
relative weights of the peaks are $\cos^4(\theta/2)$ (large peak),
$1/2\sin^2\theta = 2\sin^2(\theta/2)\cos^2(\theta/2)$ (middle
peak) and $\sin^4(\theta/2)$ (small peak) and their characteristic
velocities are $-\sigma v_a (1-\cos\theta)$,  $\sigma v_a
\cos\theta$ and $\sigma v_a (1+\cos\theta)$, respectively. The
``left'' and ''right moving'' distributions are symmetric with
respect to a $x\rightarrow -x$ reflection (see also Fig.
\ref{Fig3}). Notice that the total momentum corresponding to a
distribution described by equation (\ref{TOF_dens}) vanishes. By
analyzing the transformation (\ref{darkSts}) to the dark state
basis, we observe that $\sin\theta$ is the coefficient of the
hyperfine state $|3\rangle$. Consequently, the middle peak in the
density distribution (\ref{TOF_dens}) consists of particles in
this particular hyperfine state. The other two peaks contain
mixtures of states $|1\rangle$ and $|2\rangle$. A state-selective
measurement of particles in the hyperfine state $|3\rangle$ would
reveal a single peak structure moving to the left or to the right
with a velocity $v=v_a\cos\theta$. The dependence of the density
profile $\rho({\bf r}, t)/N$ on $\theta$ and on the ratio
$\gamma=v_a/\sqrt{\omega/m}$  for ${\bf r} = (x, 0, 0)$ and $t =
\omega^{-1}$ is shown in Fig. \ref{Fig3}.
\begin{figure}[tbp]
\begin{center}
\includegraphics[width=0.48\textwidth]{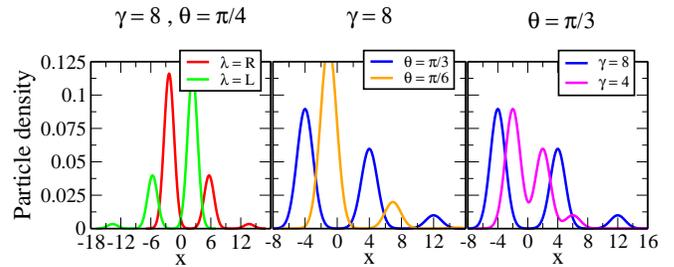}
\end{center}
\caption{(color online): Density profiles $\rho({\bf r}, t)/N$ for
${\bf r}=(x, 0, 0)$ and $t = \omega^{-1}$. The position $x$ is
measured in units of $\Gamma^{-1}$. Left panel: "right moving"
versus "left moving" distributions. Notice that the ``center of
mass'' of the distributions is always at $x=0$.  Middle panel:
Dependence on the angle $\theta$. At small angles all the weight
concentrates in the large peak which is centered near $x=0$. In
the limit $\theta\rightarrow \pi/2$ the strength of the SO
interaction vanishes $v\rightarrow 0$ and the present analysis is
not valid. Left panel: Dependence on the relative strength of the
spin-orbit coupling, $\gamma = v_a/\sqrt{\omega/m}$. To resolve
the peak structure, the spin-orbit coupling energy scale should be
larger than the trap level spacing. In the opposite limit
interference effects become important (see main text).}
\label{Fig3}
\end{figure}

\section{Summary and conclusions} \label{V}

To summarize, in this article we have introduced and discussed in detail a new type of many-body system consisting of pseudo spin-$1/2$ bosons with spin-orbit interactions.  We have shown that at low temperatures the system condenses into a new type of entagled  BEC, the spin-orbit coupled Bose-Einstein condensate (SOBEC). The novelty of this state stems from the coupling of an internal degree of freedom, the  pseudo-spin created as a result of an atom interacting with a spatially modulated laser field, to the real space motion of the particles.  As a result, the single-particle spectrum is characterized by degenerate minima at finite momenta and, consequently, the bosons condense at low temperatures into an entangled quantum state with non-zero momentum. For an  arbitrary spin-orbit coupling, the single particle spectrum has a double-well structure in momentum space (see Fig. \ref{Fig1}) with minima at non-zero momenta. In this case, a system of $N$ non-interacting bosons is characterized by a large ($N+1$)-fold degeneracy of the many-body ground state. Weak density-density interactions reduce this large degeneracy to a two-fold degeneracy. The corresponding ground state wave-function describes a superposition of left-moving and right-moving condensates with weights $w_L$ and $w_R=1 - w_L$, respectively. Performing a time-of-flight expansion of the condensed bosons results in a characteristic three-peak structure (see Fig. \ref{Fig3}). The total momentum of the density profile is identically zero, but the peaks are moving along the x-direction with velocities proportional to the k-vector of the  laser field modulation in that direction. The probability of measuring a left- (right-) moving condensate is $w_L$ ($w_R$) and the signature of a left- (right-) moving state consists in the middle and small peaks moving left (right), while the large peak moves in the opposite direction. 

In conclusion, the spin-orbit coupled BEC can be viewed as a state occurring at the interface between spintronics and cold atom physics, with nontrivial properties that have a significant potential  for applications. We note here that the ground-state of the double-well SOBEC [see Eq. (\ref{GSN})] represents a NOON  state,\cite{NOON,NOON1} similar to those recently proposed for the construction of a gravimeter bases on atom interferometry.\cite{NOON2} Therefore, the  study of a SOBEC state in the context of quantum entanglement and quantum interference is highly relevant. In addition, the double degeneracy associated with the pseudo-spin  degree of freedom makes this state a natural candidate for a qubit. A possible way to measure such a qubit was described in the last section. Time-dependent laser fields [similar to those, which lead to the spin-orbit-coupled Hamiltonian (\ref{H})] could be used as ``gates'' to perform unitary rotations in the space of degenerate ground states. Note that the coupling of the spin to the orbital motion yields a protecting mechanism against decoherence, due to momentum conservation, and suggest that the spin-orbit coupled condensates are interesting candidates for fault tolerant quantum computation. Finally, we note that for a symmetric Rashba-type spin-orbit coupling the system is characterized by a single-particle spectrum that  has a continuous set of minima along a circle in momentum space. This results in a huge degeneracy that may lead to possible phases with non-trivial topological properties, making the study of the symmetric SOBEC a potentially very interesting problem.

\vspace{2mm}

V.G. is grateful to C. Clark, I. Satija, I. Spielman, and J.
Vaishnav  for useful discussions and JQI for financial support.

\bibliography{coldspin}

\end{document}